\documentclass[12pt]{utarticle}
\usepackage{amsmath,amsfonts,bbm,mathrsfs}
\usepackage{epsfig,color}
\title{\vskip-0.5cm Embedding LFHQCD in Worldsheet String Theory}
\author{Harun Omer}
\oneaddress{UMass Dartmouth\\ 285 Old Westport Rd\\ Dartmouth MA 02747 {\tt homer@umassd.edu}}
\Abstract{Light-front holographic quantum chromodynamics (LFHQCD) has been proposed as an approximation to QCD which is non-perturbative and at the same time analytically tractable. It can be derived from the holographic light-cone Hamiltonian and effectively corresponds to a one-dimensional superconformal quantum theory. It is proposed in this paper to extend the underlying quantum group to $\mathcal{N}=2$ SCFT. This provides an avenue for bottom-up string theory model-building which is firmly grounded in phenomenology. Moreover, the peculiar properties of the model carry over to provide novel solutions to string theory problems, including supersymmetry without predicting superpartners as well as the breaking of the conformal symmetry and the introduction of a low-energy scale such that the massive Virasoro modes correspond to hadronic states reminiscent of what was once envisioned in the Veneziano model.}
\begin{document}

\maketitle 
\section{Introduction}
In the last few years, a series of papers has been published about Light-Front Holographic QCD (LFHQCD)~\cite{LFHQCD1,LFHQCD2,LFHQCD3,LFHQCD4,LFHQCD5,LFHQCD6,LFHQCD7,LFHQCD8,LFHQCD9,LFHQCD10,LFHQCD11,LFHQCD12,LFHQCD13}. Helpful reviews are~\cite{review1,review2}. LFHQCD is able to reproduce a number of spectroscopic features of hadrons within the framework of a rather simple, semi-classical non-perturbative model. One starts with anti-de Sitter space and uses the holographic principle to obtain a field theory in one lower dimension in the limit of $N_c\rightarrow \infty$ colors. The equations of motion can be matched with the bound-state equations for two massless
constituents in light front (LF) quantization. It was then realized that the resulting model can also be obtained via the  $AdS_2/CFT_1$ duality. The predicted masses for mesons $M$ and baryons $B$ are summarized as,
\begin{eqnarray}
\begin{array}{rcl}
M_M^2&=&2\lambda(n+L_M)+\lambda S\\
M_B^2&=&2\lambda(n+L_B+1)+\lambda S
\end{array}
\end{eqnarray}
where $n$ is the radial excitation quantum number, $L_M$ the orbital angular momentum between the quark and anti-quark in the meson and $L_B$ that between diquark cluster and quark in the baryon and $S$ is a spin contribution. The only free constant in the theory is the scale $\lambda$. Comparison with experimental data gives a reasonable match to typically about 10\% accuracy as is detailed in the referenced papers. Note in the above equation the shift of 1 between mesonic and baryonic masses. The underlying symmetry group of the LFHQCD model is the supersymmetric $OSp(1,2)$ with bosonic subgroup $SL(2,\mathbb{R})\simeq SO(2,1)$. This will be the starting point for this work. The conformal mechanics model based on this bosonic group was first introduced by de Alfaro, Fubini and Furlan (dAFF) in~\cite{dAFF} without focus on any particular application. Yet their work demonstrated how  a mass scale could appear in a Hamiltonian and the equations of motion, breaking dilatation invariance, all the while leaving the action conformally invariant. Their theory was extended to the supersymmetric $OSp(1,2)$ in~\cite{fubinirabinovici}. The latter reference also discusses spontaneous breaking of supersymmetry from $\mathcal{N}=2$ to $\mathcal{N}=1$. The model's application to LFHQCD has been promoted as an approximation to QCD  that is non-perturbative yet still tractable. Here a brief summary of the intriguing symmetry properties of LFHQCD:
\begin{itemize}
\item 
\textbf{Supersymmetry}: The Large Hadron Collider did not uncover evidence for superpartners, yet powerful arguments from string theory and other physics beyond the standard model point to the existence of supersymmetry for consistency reasons. It is intriguing that the model possesses susy without predicting superpartners (since susy relates meson and baryon wave functions but not fields) and despite this incorporates susy naturally, manifestly exhibiting it in the hadronic spectrum. 
\item \textbf{Conformal symmetry}: Conformal symmetry appears to be an indispensable component of string theory since it is part of the full diffeomorphism symmetry required for quantum gravity. It is also phenomenologically not far-fetched to assume the existence of conformal symmetry at some level since Maxwell's equations, which do not have any mass scales in them which could break conformal invariance, are invariant under conformal transformations. From a practical point of view, conformal invariance is also rather welcome since plenty of supersymmetric theories can be constructed whereas conformal invariance is rather restrictive and reduces their multitude to just a few. In LFHQCD conformal invariance  is not arbitrarily imposed but crucial for  its properties, specifically color confinement.
 \item \textbf{Holography}: The existence of the holographic duality is likewise attractive from a string theory perspective.
\end{itemize} 
While the model's appeal lies in its simplicity, the downside is that its simplicity makes it too rigid for refinements. For example, the model is based on the t'Hooft approximation of infinitely many colors $N_c \rightarrow \infty$. Also, along the Regge trajectories can be multiple states but the model explains only one for each combination of level, angular momentum and spin. The spin term of the model has been introduced by hand and for a full treatment of spin a 0+1 dimensional model is likely insufficient. Furthermore, the model is solved separately for each  meson flavor with slightly different parameters $\lambda$. Clearly, ultimately we want to have a single model incorporating all flavors as well as particle interactions. One approach would be to revisit QCD\ and attempt to derive more sophisticated approximations. 
This work proceeds on a different path, starting with the $OSp(1,2)$ model and embedding it into  $\mathcal{N}=2$ SCFT. Since modifications of the model are left for future work, the relevance is primarily of technical nature: The model's features lead to aspects novel to string theory, notably the way the breaking of conformal invariance by which a scale is introduced into the theory. At the same time the model is well-rooted in QCD and phenomenology and provides a basis for bottom-up string theory model-building. Following are some thoughts on why the proposed extensions are reasonable:
\begin{itemize}
\item \textbf{Diffeomorphism invariance}: The one-dimensional conformal algebra of $CFT_1$ is a subgroup of the Virasoro algebra and there are some strong indications that the algebra should be extended to the full Virasoro algebra. Generally speaking, the Virasoro algebra enhances the symmetry to reparametrization invariance, also called diffeomorphism invariance. Diffeomorphism transformations are transformations to accelerating reference frames. Newtonian mechanics clearly is not invariant under accelerating transformations, however, when classical mechanics is rewritten in a Lagragean or Hamiltonian formalism, it is covariant under accelerating transformations. Invoking Einstein's equivalence principle, accelerating transformations are equivalent to gravitational forces, which is the underlying reason why a quantum theory of gravity should possess diffeomorphism invariance. Clearly, in 0+1 dimensions it is not so clear what gravity means, but even for $CFT_1$ specifically there are compelling arguments for an enhancement to the Virasoro algebra. It was argued by Strominger in~\cite{strominger} that quantum gravity on $AdS_2$ should be a CFT on a strip and satisfy the full symmetries of the Virasoro algebra.
This led the author of \cite{kumar} to suspect that the boundary CFT should also exhibit the full Virasoro symmetries and he showed that classically a one-dimensional theory with conformal symmetry must also satisfy the larger Virasoro algebra; an analysis which was later extended under basic assumptions to the superconformal case~\cite{marcus}.
\item \textbf{Increase of the dimensionality of space:} The 0+1 dimensional model has been obtained from a higher-dimensional theory using the light-cone formalism. It is proposed to set this aside and contemplate a dual reformulation of the theory. Physically, what we have is a 0+1 dimensional theory which reproduces the essentials of the hadron spectra. Obviously we do not live in 0+1 dimensions, but there may be other ways to connect with the real world than the one taken in the original derivation. An increase in dimensionality is not as straightforward as it might na\"ively appear. The simplest conformally invariant Lagrangeans for a scalar field have the form,
\begin{eqnarray}
\mathcal{L}=\frac{1}{2}\partial_{\mu}\phi\partial^{\mu}\phi-g\phi^{\frac{2d}{d-2}}
\end{eqnarray}
The interaction term $\sim\frac{1}{\phi^{2}}$ which arises for $d=1$, is an essential feature of the model and is specific to $d=1$ only. Yet there is another way to add a spatial dimension by simply increasing the spatial dimensionality of the Virasoro algebra. We would then have a two-dimensional $\mathcal{N}=2$ Super-Virasoro algebra, also called 2d $\mathcal{N}=2$ SCFT. Note that the diffeomorphism invariance advocated for above enters automatically with an increase of dimensionality. The bosonic conformal group in $\mathbb{R}^{0,1}$ is Conf$(\mathbb{R}^{0,1})\cong SO(1,2) \subset \text{Diff}_{+}(\mathbb{R})$ but in $\mathbb{R}^{1,1}$ it is $\text{Conf}(\mathbb{R}^{1,1})\cong\text{Diff}_{+}(\mathbb{R}) \times \text{Diff}_{+}(\mathbb{R})$. The hadron physics would 'live' on a one-dimensional boundary of the theory. This is not only in line with the holographic principle, it is also a tested method in string theory, where the algebra in question appears in string compactifications down to four space-time dimensions. In string theory models, gravity propagates through closed strings in the 'bulk' of the theory (which on the holographic dual side is replaced by a curved space) and the remainder of the physics arises at the boundary. In string theory, 2d $\mathcal{N}=2$ SCFT appears as the gauge algebra of non-linear $\sigma$-models which arise in string compactifications. It also appears as the gauge algebra of the $U(1)$ string.
In string compactifications the 2d $\mathcal{N}=2$ SCFT is embedded in a higher-dimensional target space. The world-sheet boundary objects can correspond to branes. Clearly this is a radical re-interpretation of the LFHQCD model. Yet it provides a chance to extend it to a more realistic model. For instance a stack of $N_c=3$ branes gives rise to a $SU(3)$ gauge symmetry. While the framework of string theory is used, it is nevertheless a radical departure from what has been done in the field in the last decades and in a sense a return to the beginnings. String theory started with the Veneziano model intended to reproduce Regge trajectories for QCD and it was realized it corresponded to a theory of strings. When it turned out that it suffered from flaws,  string theory moved towards alternative approaches.  
\end{itemize}
\section{Extension from the $OSp(1,2)$ to the Virasoro Symmetry}

The $\mathcal{N}=2$ Super-Virasoro algebra, also called the $\mathcal{N}=2$ superconformal algebra, is defined by the brackets (see e.g.\cite{schomerusprimer}),
\begin{eqnarray}
\begin{array}{rcl}
\,[L_m,L_n] &=& (m - n)L_{n+m} + \frac{C}{12}(m^3-m)\delta_{m+n,0}\\
\;[L_m,G_r^{\pm}]&=&\left(\frac{1}{2}m-r\right)G^{\pm}_{m+r}\\
\;[L_m,J_n]&=&-nJ_{m+n}\\
\;\{G^{+}_r,G^{-}_s\}&=&2L_{r+s}+(r-s)J_{r+s}+\frac{C}{3}\left(r^2-\frac{1}{4}\right)\delta_{r+s,0}\\
\;[J_n,G^{\pm}_r]&=&\pm G^{\pm}_{n+r}\\
\;[J_m,J_n]&=&\frac{C}{3}m\delta_{m+n,0}
\end{array}
\end{eqnarray}
Here $C$ is the central charge. The generators $L_n$ correspond to the modes of the stress-energy tensor $T$, the $J_n$ are the modes of  the $U(1)$-current $J$ and the fermionic generators $G_{r}^{\pm}$ are the modes of the two fermionic currents $G^{\pm}$.
The indices of $G^{\pm}$ either take only integer values (Ramond algebra) or only half-integer values (Neveu-Schwarz algebra). In two dimensions, one has a second copy of the Virasoro algebra with generators typically denoted as $\bar{L}_{m}$. The spectral shift isomorphism maps the two isomorphic cases into one another. Apart from these two versions of the algebra, there are also two topologically twisted versions of it. 
The $\mathcal{N}=2$ Superconformal algebra contains the algebra of conformal quantum mechanics as a finite dimensional subalgebra. In the Neveu-Schwarz (NS) sector one identifies,
\begin{eqnarray}
\displaystyle \{H,iD,-K,Q,Q^{\dagger},-S,S^{\dagger},B-f\mathbb{I}\}:=\{ L'_{-1}, L'_0, L'_1,G'^{{+}}_{-\frac{1}{2}},G'^{{-}}_{-\frac{1}{2}}, G'^{{+}}_{\frac{1}{2}}, G'^{{-}}_{\frac{1}{2}},J'_{0}\},\label{eq:subgr}
\end{eqnarray}
where a prime was added to the Virasoro operators. Here $H$ is the translation operator $D$ the generator of dilatations and $K$ the generator of conformal transformations. The complete list of non-trivial brackets is as follows:
\begin{eqnarray}
\begin{array}{rcllll}
\,[L'_{-1},L'_0] &=& -L'_{-1} &&[H,D]=iH\\
\,[L'_{-1},L'_1] &=& -2 L'_{0} && [H,K]=2iD\\
\,[L'_1,L'_0] &=& L'_{1} && [K,D]=-iK\\
\;[L'_{-1},G'^{\pm}_{\frac{1}{2}}]&=&-G'^{\pm}_{-{\frac{1}{2}}} && [H,S]=Q & [H,S^{\dagger}]=-Q^{\dagger} \\
\;[L'_0,G'^{\pm}_{\frac{1}{2}}{}]&=&-\frac{1}{2}G'^{\pm}_{\frac{1}{2}} && [D,S]=\frac{i}{2}S & [D,S^{\dagger}]=\frac{i}{2}S^{\dagger}\\
\;[L'_0,G'^{\pm}_{-\frac{1}{2}}]&=&\frac{1}{2}G'^{\pm}_{-\frac{1}{2}}&& [D,Q]=-\frac{i}{2}Q & [D,Q^{\dagger}]=-\frac{i}{2}Q^{\dagger}\\
\;[L'_1,G'^{\pm}_{-\frac{1}{2}}]&=&G'^{\pm}_{\frac{1}{2}}&& [K,Q]=S& [K,Q^{\dagger}]=-S^{\dagger}\\
\;\{ G'^{+}_{\frac{1}{2}},G'^{-}_{\frac{1}{2}}\}&=&2 L'_{1} &\;\;\;\;\longmapsto\;\;\;\;& \{S,S^{\dagger}\}=2K\\
\;\{G'^{+}_{\frac{1}{2}},G'^{-}_{-\frac{1}{2}}\}&=&2L'_{0}+J'_{0}&& \{S,Q^{\dagger}\}=-2iD+f\mathbb{I}-B\\
\;\{G'^{+}_{-\frac{1}{2}},G'^{-}_{\frac{1}{2}}\}&=&2 L'_{0}-J'_{0}&&\{Q,S^{\dagger}\}=2iD+f\mathbb{I}-B\\
\;\{G'^{+}_{-\frac{1}{2}},G'^{-}_{-\frac{1}{2}}\}&=&2L_{-1}' && \{Q,Q^{\dagger}\}=2H\\
\;[J'_0,G'^{+}_{\frac{1}{2}}]&=&G'^{+}_{\frac{1}{2}} && [B,S]=S\\
\;[J'_0,G'^{-}_{\frac{1}{2}}]&=&-G'^{-}_{\frac{1}{2}} && [B,S^{\dagger}]=-S^{\dagger}\\
\;[J'_0,G'^{+}_{-\frac{1}{2}}]&=& G'^{+}_{-\frac{1}{2}} && [B,Q]=Q\\
\;[J'_0,G'^{-}_{-\frac{1}{2}}]&=&-G'^{-}_{-\frac{1}{2}} && [B,Q^{\dagger}]=-Q^{\dagger}\\
\end{array}
\end{eqnarray}
While this is a perfectly valid realization of the subalgebra, it is not in agreement with the behavior under adjungation familiar from the Virasoro algebra when it is obtained from radial quantization. Although in one dimension the adjoint operation does not apply in the usual sense, even there it makes sense to choose a unitary representation where it formally holds. In the superconformal algebra the bosonic operators are self-adjoint,
\begin{eqnarray}
H^{\dagger}=H \qquad K^{\dagger}=K \qquad D^{\dagger}=D \qquad B^{\dagger}=B
\end{eqnarray}
and the fermionic $Q$ and $Q^{\dagger}$ as well as $S$ and $S^{\dagger}$ are adjoints of each other respectively, as notation already indicates. 
On the other hand, the adjungation acts on the Virasoro  operators according to,
\begin{eqnarray}
 L_m^{\dagger}=L_{-m} \qquad J_m^{\dagger}=J_{-m} \qquad (G^{\pm}_r)^{\dagger}= G_{-r}^{\mp}.\label{eq:adjbeh}
\end{eqnarray}
 A realization of the subgroup with this desired behavior is given by,
\begin{eqnarray}
\begin{array}{rccrr}
H&=&L'_{-1}&=&L_0-\frac{1}{2}(L_{-1}+L_1)\\
iD&=&L'_{0}&=&\frac{1}{2}(L_{-1}-L_1)\\
-K&=&{L'}_1&=&-L_0-\frac{1}{2}(L_{-1}+L_1)\\
Q&=&G'^{+}_{-\frac{1}{2}}&=&\frac{1}{\sqrt{2}}(G^{+}_{-\frac{1}{2}}-G^{+}_{\frac{1}{2}})\\
Q^{\dagger}&=&G'^{-}_{-\frac{1}{2}}&=&-\frac{1}{\sqrt{2}}(G^{-}_{-\frac{1}{2}}-G^{-}_{\frac{1}{2}})\\
-S&=&G'^{+}_{\frac{1}{2}}&=&-\frac{1}{\sqrt{2}}(G^{+}_{-\frac{1}{2}}+G^{+}_{\frac{1}{2}})\\
S^{\dagger}&=&G'^{-}_{\frac{1}{2}}&=&\frac{1}{\sqrt{2}}(G^{-}_{-\frac{1}{2}}+G^{-}_{\frac{1}{2}})\\
B-f\mathbb{I}&=&J'_0&=&J_0
\end{array}\label{eq:subalgebra}
\end{eqnarray}
The mapping between the primed and unprimed representation is its own inverse.
Note that the authors of~\cite{dAFF} arrived at the same transformation when solving their model.
\section{Two-dimensional world-sheets}
We pause to review some known facts about boundary conformal field theory so that it will become clear that the proposed extension is compatible with the usual boundary conditions in SCFT. Readers familiar with boundary SCFT may skip this section. In string theory the two-dimensional SCFT is a world-sheet theory with space-time indices attached to the generators. These have been suppressed so far but can be reinstated easily. Of particular importance in string compactification are superconformal field theories with $\mathcal{N}=(2,2)$ supersymmetry. The notation indicates  supersymmetry in both the holomorphic (left-moving) and the anti-holomorphic (right-moving) sector. From a world-sheet perspective, $D$-branes arise as via a boundary condition imposed on the world-sheet. Typically one works in complex coordinates and restricts the domain to the upper half plane. Along the boundary, consistency conditions between the holomorphic and anti-holomorphic currents have to be imposed. For instance on the boundary $z=\bar{z}$ the holomorphic and anti-holomorphic component of the stress energy tensor have to coincide, \begin{eqnarray}
T(z)=\bar{T}(\bar{z}).
\end{eqnarray}
The physical interpretation of the constraint is to prevent energy from flowing through the boundary. For the modes of the stress-energy tensor this implies $\bar{L}_n=L_{-n}$.
For the other modes there are two choices of boundary conditions,
\begin{eqnarray}
\begin{array}{lll}
\text{A-type: } & J(z)=-\bar{J}(\bar{z}) &\ G^{\pm}(z)=\eta \bar{G}^{\mp}(\bar{z})\\
\text{B-type: } &J(z)= \bar{J}(\bar{z}) &\ G^{\pm}(z)=\eta \bar{G}^{\pm}(\bar{z})
\end{array}\qquad \eta = \pm 1.
\end{eqnarray}
The two are related by the mirror automorphism $J\mapsto -J, G^{\pm}\mapsto G^{\mp}$. The gluing conditions reduce the $\mathcal{N}=(2,2)$\ supersymmetry to ordinary two-dimensional supersymmetry. Since the structure of the anti-holomorphic sector with generators $\bar{L}_n$, $\bar{G}^{\pm}_r$, $\bar{J}_n$ is parallel to the holomorphic sector and the two are related through the gluing conditions, it generally suffices to work with only the holomorphic generators.
A good reference on world-sheet string theory is~\cite{cardy}.
\section{Algebraic Solution of the Quantum Theory}
The Virasoro algebra is dealt with in every textbook of string theory such as in the classic~\cite{gsw}. The (super-)conformal model is solved in the original papers~\cite{dAFF,fubinirabinovici}. This section serves to clearly establish the connection to the Virasoro algebra. In the original literature the (super)-conformal model is derived from Lagrangeans, which is not helpful for the purpose of this work. Here a purely algebraic procedure is followed. To find the unitary irreducible representations of the group one first needs to find a complete set of commuting observables (CSCO). The eigenstates are then labeled by their eigenvalues, although labels are often suppressed for simplicity. We consider a unitary irreducible highest weight representation which has a normalized highest weight state $|r_{0},q,c\rangle$ and is defined by,
\begin{align}
\begin{array}{rcll}
L_0|r_{0},q,c\rangle&=&r_{0}|r_{0},q,c\rangle\\
J_0|r_{0},q,c\rangle&=&q|r_{0},q,c\rangle\\
C|r_{0},q,c\rangle&=&c|r_{0},q,c\rangle\\
L_n|r_{0},q,c\rangle&=&0 &\forall \;n>0 \\
G^{\pm}_r|r_{0},q,c\rangle&=&0 & \forall \;r>0 \\
J_n|r_0,q,c\rangle&=&0 &\forall \;n>0
\end{array}
\end{align}
Unitarity implies $L_n^{\dagger}=L_{-n}$ for all generators. The operators $L_n$ are ladder operators which change the eigenvalue of $L_0$ by $n$,
\begin{eqnarray}
\;[L_0,L_n]=-nL_n.
\end{eqnarray}
Note that in order for a lowest weight state to exist, the ladder property requires the condition $L_n|r_0,q,c\rangle=0$ for $n>0$.
According to the Poincar\'e-Birkhoff-Witt theorem, all states are of the form,
\begin{eqnarray}
\begin{array}{rcl}
|\phi\rangle&=&(L_{-k})^{n_k}\dots (L_{-2})^{n_2}(L_{-1})^{n_1}|r_{0},q,c\rangle\\
L_0|\phi\rangle&=&(r_{0}+N)|\phi\rangle=(r_{0}+kn_k+(k-1)n_{k-1}+...2n_2+n_1)|\phi\rangle
\end{array}\qquad n_j,N\in \mathbb{N}\label{eq:level}
\end{eqnarray}
$N$ is called the level of the state. The normalization of  states is found using,
\begin{eqnarray}
||L_{-n}|r_{0},q,c\rangle||^2=\langle r_{0},q,c|L_nL_{-n}|r_{0},q,c\rangle=2nr_{0}+\frac{c}{12}n(n^2-1)\qquad n\in \mathbb{N}_0 \label{eq:norm1}
\end{eqnarray}
It is easy to see that a positive semi-definite norm implies $r_{0},c\ge 0$. In the $SO(2,1)$ theory, the normalized eigenstates are proportional to $(L_{-1})^n|r_{0},q,c\rangle$. Their normalization is found from,
\begin{eqnarray}
||(L_{-1})^n|r_{0},q,c\rangle||^2=\langle r_0,c,q|(L_{1})^n(L_{-1})^n|r_{0},q,c\rangle
\end{eqnarray}
and then iteratively applying $L_1L_{-1} = 2L_{0}+L_{-1}L_1$ together with Eq.~(\ref{eq:level}). The result is,
\begin{eqnarray}
\begin{array}{rcl}
(L_{-1})^n|r_{0},q,c\rangle&=&\sqrt{n(2r_0+n-1)}(L_{-1})^{n-1}|r_{0},q,c\rangle\\
&=&\sqrt{(n+r_0)(n+r_0-1)-r_0(r_0-1)}(L_{-1})^{n-1}|r_{0},q,c\rangle.\label{eq:norm2}
\end{array}
\end{eqnarray}
Using a similar procedure it is established that the states are orthogonal. This normalization as well as the eigenvalues $r_0+n$ of,
\begin{eqnarray}
L_0(L_{-1})^n|r_{0},q,c\rangle=(n+r_0)(L_{-1})^n|r_{0},q,c\rangle\qquad n\in \mathbb{N}_0,
\label{eq:bosev}
\end{eqnarray}
match the results of~\cite{dAFF}. In treatises on the bosonic string and the $\mathcal{N}=1$ superstring, the lowest energy eigenvalue $r_0$ is taken to be $0$ in which case the tower of $L_{-1}$ descendant states all have zero norm as can be seen from Eq.~(\ref{eq:norm1}) and Eq.~(\ref{eq:norm2}). They are thus considered pure gauge and disregarded. As an extension to (super-)conformal mechanics,  $r_0$ is generally not zero since it is related by $r_0=\frac{1}{2}(1\pm\sqrt{g+\frac{1}{4}})$ to the coupling $g$ in the conformal Hamiltonian~\cite{fubinirabinovici},
\begin{eqnarray}
H=\frac{1}{2}\left(p^2-\frac{g}{x^2}\right).
\end{eqnarray}
New as the result of the Virasoro enlargement is the existence of degenerate eigenstates for levels $n \ge 2$. At $n=2,$ two distinct states $(L_{-1})^2|r_0,q,c\rangle$ and $L_{-2}|r_0,q,c\rangle$ exist. To verify that higher-level states are true physical states, one has to remove the ones which can be expressed as linear combinations of others at the same level as well as any states with negative norm. This analysis can be performed using the Kac-determinant and the result depends on the values of $r_0$ and the central charge $c$. 
\section{The Conformal Scaling}
The operator $D$ effects a conformal rescaling by a constant factor: 
\begin{eqnarray}
\begin{array}{rclrcl}
e^{\mathrm{i}\omega D} H e^{-\mathrm{i}\omega D} &=& He^{-\omega}&=&\frac{1}{\lambda}H\\
e^{\mathrm{i}\omega D} D e^{-\mathrm{i}\omega D} &=& D\\
e^{\mathrm{i}\omega D} K e^{-\mathrm{i}\omega D} &=& Ke^{\omega}&=&\lambda K\\
e^{\mathrm{i}\omega D} B e^{-\mathrm{i}\omega D} &=& B
\end{array}\qquad \omega:=\ln \lambda \in \mathbb{R}.\label{eq:confscal}
\end{eqnarray}
This is obvious from the conformal scaling dimensions of the operators but it can also be easily verified by direct computation. These identities are needed further below to solve the LFHQCD model. For that purpose note that,
\begin{eqnarray}
U_{D}(\lambda):=\exp\{\mathrm{iD}\ln \lambda\}=\exp\{L'_0\ln \lambda\}=\exp\{{\scriptstyle \frac{1}{2}}(L_{-1}-L_1)\ln\lambda\}.\label{eq:confscal2}
\end{eqnarray}
The conformal rescaling causes $H$ to have a discrete eigenvalue spectrum; if $|\Psi\rangle$ is an eigenstate of $H$ with eigenvalue $E$, then $e^{\mathrm{i\omega D}}|\Psi\rangle$ is eigenstate with eigenvalue $e^{-\omega}E$. This motivated aDFF to replace $H$ with a more well-behaved Hamiltonian with normalizable eigenstates and they arrived at the bosonic linear combinations in Eq.~(\ref{eq:subalgebra}).
As a unitary operator, the action of $U_D(\lambda)$ on all operators of the symmetry algebra is an automorphism of the algebra. Since the transformed theory is unitarily equivalent to the free theory, one can always transform it back to the free theory and might therefore just as well ignore the action of $U_D(\lambda)$. Yet keep in mind that this only holds as long as $\lambda$ is a constant, global parameter. If it is elevated to a dynamic, local variable, it can not be gauged away anymore. It is worth remarking that $\lambda$ is a dimensionful parameter; as is evident from Eq.~(\ref{eq:subalgebra}), the presence of $\lambda$ ensures that the dimensions of $\frac{1}{\lambda}H$ and $\lambda K$ match. However, it is not $U_D(\lambda)$ which is responsible for the appearance of the scale in the theory. No unitary operator can achieve that. Take a quantum system with eigenvalue equation $H |\Psi\rangle = E|\Psi\rangle$ and transform this system unitarily to  $H_U=UHU^{-1}$ and $|\Psi'\rangle = U |\Psi \rangle$. The new system $H_U|\Psi'\rangle=E_U|\Psi'\rangle$ has an identical eigenvalue $E_U=E$. If the characteristic energy scale of the theory is at the Planck scale, as would be expected for any theory of quantum gravity, no unitary transformation is able to change this scale.  
\section{The LFHQCD model}
In the LFHQCD model, the charge $Q$ is replaced with $R_{\lambda}=Q+\lambda S$. It is this modification through which the scale $\lambda$ enters the model. The Hamiltonian acquires two additional terms:
\begin{eqnarray}
H_{\lambda}=\frac{1}{2}\{R_{\lambda},R_{\lambda}^{\dagger}\}=H+\lambda(f- B)+\lambda^2 K.
\end{eqnarray}
In the Virasoro notation the redefined supercharges read,
\begin{eqnarray}
\begin{array}{rclrclrcl}
R^{+}_{\lambda}&=&Q+\lambda S&=&{G'}^{+}_{-\frac{1}{2}}-\lambda G'^{+}_{\frac{1}{2}}&=&\frac{1}{\sqrt{2}}G^{+}_{-\frac{1}{2}}(1+\lambda)-\frac{1}{\sqrt{2}}G^{+}_{\frac{1}{2}}(1-\lambda)\\
R^{-}_{\lambda}&=&Q^{\dagger}+\lambda S^{\dagger}&=&G'^{-}_{-\frac{1}{2}} + \lambda G'^{-}_{\frac{1}{2}}&=&\frac{1}{\sqrt{2}}G^{-}_{\frac{1}{2}}(1+\lambda)-\frac{1}{\sqrt{2}}G^{-}_{-\frac{1}{2}}(1-\lambda).
\end{array}
\end{eqnarray}
This redefinition can be generated by
a similarity transformation with the non-unitary operator $V=e^{\lambda K}=e^{-\lambda L'_1}=e^{\lambda(L_0+\frac{1}{2}(L_{-1}+L_1))}$,
\begin{eqnarray}
\begin{array}{rcll}
VG'^{+}_{-\frac{1}{2}}V^{-1}&=&G'^{+}_{-\frac{1}{2}}-\lambda G'^{+}_{\frac{1}{2}} &  (\text{or }VQV^{-1} = Q+\lambda S),\\
VG'^{+}_{\frac{1}{2}}V^{-1}&=&G'^{+}_{\frac{1}{2}}\qquad  &(\text{or }VSV^{-1} =  S).
\end{array}
\end{eqnarray}
From the new supercharges one obtains,
\begin{eqnarray}
\begin{array}{rcl}
H_{\lambda}=\frac{1}{2}\{R^{+}_{\lambda},R^{-}_{\lambda}\}
&=&L'_{-1}-\lambda J'_{0}- \lambda^{2}L'_{1}\\
&=&L_0(1+\lambda^2)-\lambda J_{0} - \frac{1}{2}(1-\lambda^2)(L_{-1}+L_1)\\
\{R^{+}_{\lambda},R^{+}_{\lambda}\}&=&0\\
\{R^{-}_{\lambda},R^{-}_{\lambda}\}&=&0
 \end{array}
\end{eqnarray}
The supersymmetry structure remains unaffected.
The Hamiltonian still looks somewhat complicated, but through a conformal rescaling using Eqs.~(\ref{eq:confscal}) and Eqs.~(\ref{eq:confscal2}), the different powers of the scaling parameter $\lambda$ are consolidated into one overall scaling factor:
\begin{eqnarray}
\boxed{
\begin{array}{rcl}
H_{LFHQCD}:=U_{D}(\lambda)^{-1}H_{\lambda}U_D(\lambda)
&=&\lambda(L'_{-1}- J'_{0}- L'_1)\\
&=&\lambda\{G^{+}_{-\frac{1}{2}},G^{-}_{\frac{1}{2}}\}\\
&=&\lambda(2L_0- J_{0})
\end{array}
}
\label{eq:finalham}
\end{eqnarray}
The result is rather familiar from $\mathcal{N}=2$ SCFT except for the striking scaling factor $\lambda$. Note that any eigenstate $|\psi\rangle$ has the same eigenvalue as an arbitrary linear combination of $G^{+}_{-\frac{1}{2}}|\psi\rangle$ and $G^{-}_{\frac{1}{2}}|\psi\rangle$ since both $G^{+}_{-\frac{1}{2}}$ and $G^{-}_{\frac{1}{2}}$ commute with $2L_0-J_0$.
Combining Eq.~(\ref{eq:bosev}) and~(\ref{eq:finalham}) gives,
\begin{eqnarray}
\boxed{H_{LFHQCD}(L_{-1})^n|r_{0},q,c\rangle=2\lambda(n+r_0-{\scriptstyle \frac{1}{2}}q)(L_{-1})^n|r_{0},q,c\rangle\qquad n\in \mathbb{N}_0.}\label{eq:lfhqcdeqn}
\end{eqnarray}
Recall that chiral and anti-chiral states in the NS-sector are defined by,
\begin{eqnarray}
\begin{array}{lr}
\text{chiral: }&G^{+}_{-\frac{1}{2}}|\phi\rangle=0\\
\text{anti-chiral: }&G^{-}_{-\frac{1}{2}}|\phi\rangle=0
\end{array}
\end{eqnarray}
Chiral primary states are the states for which the eigenvalue of our Hamiltonian vanishes,
\begin{eqnarray}
\{G^{+}_{-\frac{1}{2}},G^{-}_{\frac{1}{2}}\}|\phi\rangle=0
\end{eqnarray}
For a chiral primary $r_0=2q$. It can be shown (see e.g.~\cite{chiralrings}) that any state can be written in the form,
\begin{eqnarray}
|\psi\rangle = |\phi\rangle + G^{+}_{-\frac{1}{2}}|\chi_1\rangle + G^{-}_{\frac{1}{2}}|\chi_2\rangle,
\end{eqnarray}
where $|\phi\rangle$ is a chiral primary and $|\chi_{1,2}\rangle$ are other states. There are a  number of constraints from unitarity such as,
\begin{eqnarray}
\begin{array}{rcl}
||G^{\pm}_{-\frac{1}{2}}|r_0,q,c\rangle||^2\ge 0 &\Rightarrow& r_0 \ge \frac{|q|}{2}\\
||G^{\pm}_{-\frac{3}{2}}|r_0,q,c\rangle||^2\ge 0 &\Rightarrow& 2r_0\pm 3q+\frac{2}{3}c \ge 0
\end{array}
\end{eqnarray}
The bound of the first constraint is saturated by the chiral primary states. Such constraints have been discussed in detail in~\cite{boucheretal}. Let us proceed to look at the parameters of Eq.~(\ref{eq:lfhqcdeqn}) in detail. The eigenstates of the $U(1)$ generator $B$ in the superconformal model of~\cite{fubinirabinovici} were $\pm \frac{1}{2}$. Since $J_0=B-f\mathbb{I}$, for  $f= 0$ the charge eigenvalues are  $q=\pm\frac{1}{2}$. Associated with each sign  is a different highest weight state. The unitarity bound leads to the limit $r_0\ge \frac{1}{4}$ for both states. The same bound was found in the models~of \cite{dAFF,fubinirabinovici}. We may shift the $U(1)$ charge by a constant $f$ so that $q=-f\pm\frac{1}{2}$. Then,
\begin{eqnarray}
H_{LFHQCD}G^{+}_{-\frac{1}{2}}(L_{-1})^{n}|r_0,-f\pm{\textstyle \frac{1}{2}},c\rangle=2\lambda(r_{0}+n+{\textstyle \frac{1}{2}}f\pm{\textstyle\frac{1}{4}})G^{+}_{-\frac{1}{2}}(L_{-1})^{n}|r_0,-f\pm{\textstyle\frac{1}{2}},c\rangle.
\end{eqnarray}
Of course other choices for the parameters are possible. In the LFHQCD\ model, $r_0$ is related to the hadronic angular momentum $L$.  In~\cite{LFHQCD1} it has been set to $r_0=L+1$ for mesons. It originates from the duality between AdS and the light-front quantization. 
\section{Summary}
In this paper the symmetry group of light-front holographic QCD was extended to the group of $\mathcal{ N}=2$ SCFT familiar from string compactifications and it was shown how the solution to LFHQCD looks in this extended framework. The steps followed in the derivation can of course be restricted to the finite subgroup to recover the familiar results with just a renaming of operators. The derivation of this work differs from the quoted references in that it is purely algebraic, which lends itself better to the generalization of the larger group. In the references the Hamiltonian is defined  as $G=uH+vD+wK$ where the condition that the operator $G$ remain compact imposes a condition on the three real parameters $u,v,w$. In this work a unitary transformation $U=e^{i \omega D}$ has been applied to eliminate a redundancy in these parameters and isolate the overall scaling factor. The main features of the LFHQCD\ model as reviewed in the introduction carry over into string theory, providing novel avenues to string theory problems. Most notable is the introduction of the scale so that the tower of eigenstates no longer have energies on the order of the Planck scale and the lowest state is not necessarily of zero energy. While there are parallels to the original Veneziano model of string theory, it is not identical. Extensions to the LFHQCD model itself may proceed in different directions and are therefore not considered herein. 
\section{Acknowledgements}
I would like to thank Stanley Brodsky and Guy de T\'eramond for helpful discussions and comments.


\begin{thebibliography}{99}
\bibitem{LFHQCD1} G. F. de T\'eramond, H. G. Dosch, S. J. Brodsky, {\it "Baryon spectrum from superconformal quantum mechanics and its light-front holographic embedding"}, Phys. Rev. D \textbf{91}, 045040 (2015) {\tt [arXiv:1411.5243
[hep-ph]]}.
\bibitem{LFHQCD2}  H. G. Dosch, G. F. de T\'eramond, S. J. Brodsky, {\it "Superconformal baryon-meson symmetry and light-front holographic QCD}, Phys. Rev. D \textbf{91}, 085016 (2015) {\tt [arXiv:1501.00959 [hep-th]]}.
\bibitem{LFHQCD3} H. G. Dosch, G. F. de T\'eramond, S. J. Brodsky, {\it "Supersymmetry across the light and heavy-light hadronic spectrum"}, Phys. Rev. D \textbf{92}, 074010 (2015) {\tt [arXiv:1504.05112 [hep-ph]]}.
\bibitem{LFHQCD4} S. J. Brodsky, G. F. de T\'eramond, H. G. Dosch, C. Lorc\'e, {\it "Universal effective hadron dynamics
from superconformal algebra"}, Phys. Lett. B \textbf{759}, 171 (2016) {\tt [arXiv:1604.06746 [hep-ph]]}.
\bibitem{LFHQCD5} H. G. Dosch, G. F. de T\'eramond, S. J. Brodsky, {\it "Supersymmetry across the light and heavy-light hadronic spectrum II"}, Phys. Rev. D \textbf{95}, 034016 (2017) {\tt [arXiv:1612.02370 [hep-ph]]}.
\bibitem{LFHQCD6} M. Nielsen, S. J. Brodsky, {\it "Hadronic superpartners from a superconformal and supersymmetric algebra"}, Phys. Rev. D 97, 114001 (2018) {\tt [arXiv:1802.09652 [hep-ph]]}.
\bibitem{LFHQCD7}  M. Nielsen, S. J. Brodsky, G. F. de T\'eramond, H. G. Dosch, F. S. Navarra, L. Zou, {\it "Supersymmetry in
the double-heavy hadronic spectrum"}, Phys. Rev. D 98, 034002 (2018) {\tt [arXiv:1805.11567 [hep-ph]]}.
\bibitem{LFHQCD8} S. J. Brodsky, G. F. de T\'eramond, {\it "Light-front hadron dynamics and AdS/CFT correspondence"},
Phys. Lett. B 582, 211 (2004) {\tt [arXiv:0310227 [hep-th]]}.
\bibitem{LFHQCD9} G. F. de T\'eramond, S. J. Brodsky, {\it "Light-front holography: A first approximation to QCD"}, Phys.
Rev. Lett. 102, 081601 (2009) {\tt [arXiv:0809.4899 [hep-ph]]}.
\bibitem{LFHQCD10} S. J. Brodsky, G. F. de T\'eramond, H. G. Dosch, J. Erlich, {\it "Light-front holographic QCD and emerging confinement"}, Phys. Rep. 584, 1 (2015) {\tt [arXiv:1407.8131 [hep-ph]]}.
\bibitem{LFHQCD11} G. F. de T\'eramond, T. Liu, R. S. Sufian, H. G. Dosch, S. J. Brodsky, A. Deur, {\it "Universality of generalized parton distributions in light-front holographic QCD"}, Phys. Rev. Lett. 120 182001 (2018)
{\tt [arXiv:1801.09154 [hep-ph]]}.
\bibitem{LFHQCD12} R. S. Sufian, T. Liu, G. F. de T\'eramond, H. G. Dosch, S. J. Brodsky, A. Deur, M. T. Islam, B. Q. Ma, {\it "Nonperturbative strange-quark sea from lattice QCD, light-front holography, and mesonbaryon fluctuation models"}, Phys. Rev. D 98, 114004 (2018) {\tt [arXiv:1809.04975 [hep-ph]]}.
\bibitem{LFHQCD13} S.J.~Brodsky, A.~Deur, {\it "Hadron and Nuclear Physics on the Light Front"}, {\tt arXiv:1809.01043 [hep-ph]]}.
\bibitem{review1} S.J.~Brodsky, {\it "Hadron Spectroscopy and Dynamics from Light-Front Holography and
Superconformal Algebra"}, {\tt arXiv:1802.08552 [hep-ph]]}.
\bibitem{review2} L.~Zhou, H.G.~Dosch, {\it "A very Practical Guide to Light Front
Holographic QCD"}, {\tt arXiv:1801.00607 [hep-ph]]}.
\bibitem{dAFF} V. de Alfaro, S. Fubini, G. Furlan, {\it "Conformal invariance in quantum mechanics"}, Nuovo Cim. A 34, 569 (1976).
\bibitem{fubinirabinovici} S.~Fubini, E.~Rabinovici, {\it "Superconformal quantum mechanics"}, Nucl. Phys. \textbf{B} 245, 17 (1984).
\bibitem{schomerusprimer} V.~Schomerus, {\it "A Primer on String Theory}", Cambridge University Press (2017).
\bibitem{strominger} A.~Strominger, {\it "$AdS_2$ Quantum Gravity and String Theory}, JHEP 01 (1999) 007 {\tt [arXiv:9809027 [hep-th]]}.
\bibitem{kumar} J.~Kumar, {\it "Conformal Mechanics and the Virasoro Algebra"}, JHEP 04(1999) 006 {\tt [arXiv:9901139 [hep-th]]}.
\bibitem{marcus} A.~Marcus, {\it "Superconformal Mechanics and the Super Virasoro Algebra"}, JHEP 0102 (2001) 043 {\tt [arXiv:0101017 [hep-th]]}.
\bibitem{gsw} M.~Green, J.~Schwarz, E.~Witten, {\it "Superstring theory Volume 1 and 2"}, Cambridge University Press (1988).
\bibitem{chiralrings} W.~Lerche, C.~Vafa, N.P.~Warner, {\it "Chiral rings in N = 2 superconformal theories"}, Nucl. Phys. \textbf{B} 324 (1989) 427.
\bibitem{boucheretal} W.~Boucher, D.~Friedan, A.~Kent, {\it "Determinant Formulae And Unitarity For The N=2 Superconformal Algebras In Two-Dimensions Or Exact Results On String Compactification"},
Phys. Lett. \textbf{B} 172 (1986) 316.
\bibitem{cardy} J.~Cardy, {\it "Boundary conformal field theory"}, {\tt [arXiv:0411189 [hep-th]]}.
\bibitem{rs_book} A.~Recknagel, V.~Schomerus, {\it "Boundary Conformal Field Theory and the Worldsheet Approach to D-Branes"}, Cambridge University Press (2013).
\end{thebibliography}
\end{document}